\documentclass[9pt, conference]{IEEEtran}
\IEEEoverridecommandlockouts
\usepackage{cite}
\usepackage{amsmath,amssymb,amsfonts}
\usepackage{graphicx}
\usepackage{textcomp}

\newcommand*{\rom}[1]{\expandafter\@slowromancap\romannumeral #1@}
\makeatother
\usepackage{algorithm, algorithmic}
\usepackage{mathtools}
\usepackage{amsmath}
\usepackage{amssymb}
\usepackage{cite}
\usepackage{threeparttable}
\usepackage{calc}
\usepackage[table]{xcolor}
\newcommand{\mycc}{\cellcolor{lightgray}}
\usepackage{textgreek}
\usepackage{booktabs}
\usepackage{siunitx}
\usepackage{array}
\usepackage{balance}

\newcolumntype{L}{>{\phantom{$\mathbin{-}$}$}l<{$}}

\newcolumntype{M}[1]{>{\centering\arraybackslash}m{#1}}
\newcolumntype{N}{@{}m{0pt}@{}}
\usepackage{url}
\usepackage{hyperref}
\usepackage{makecell}
\usepackage{enumitem}
\usepackage{ragged2e} 
\usepackage{multirow}

\def\BibTeX{{\rm B\kern-.05em{\sc i\kern-.025em b}\kern-.08em
    T\kern-.1667em\lower.7ex\hbox{E}\kern-.125emX}}
\begin{document}

\title{ORACLE: A Multi-\underline{O}bjective \underline{R}einforcement Learning-Based \underline{A}nalog \underline{C}ircuit Design Optimizer with \underline{L}arge Language Models-Guided \underline{E}xploration}

\author{\IEEEauthorblockN{Osei~Brempong, Mohammed~Ayman~Habib, Vivan~Poddar, Morteza~Fayazi}
Email: \{osei.b, m.habib, vivan.poddar, fayazi\}@utah.edu
\IEEEauthorblockA{\textit{University of Utah, Salt Lake City, UT, USA}}}

\maketitle
\begin{abstract}
Analog circuit design automation using reinforcement learning (RL) has emerged as a promising approach for reducing manual effort. However, many existing RL-based methods focus on single-objective optimization. Even methods designed for multi-objective (MO) problems often reduce multiple design specifications to a single scalar reward. This simplification limits the ability to capture the true Pareto trade-off among competing objectives and often leads to suboptimal designs. Moreover, requiring the model to be retrained from scratch whenever the desired MO specifications change remains a key limitation. To address these challenges, we present ORACLE, an open-source RL-based framework for MO analog circuit design optimization that replaces scalar reward optimization with vector-valued learning and preference-aware conditioning. 
ORACLE represents a true MO analog circuit design optimizer that uses a preference vector to specify the relative weights of multiple objectives, enabling a single trained model to generate designs across diverse trade-off settings without retraining. We further propose two preference-guidance strategies, namely normalized-weight guidance and cosine-aligned guidance, to improve convergence. In addition, we incorporate a large language model (LLM)-guided action selection mechanism to filter actions that are likely to lead to suboptimal designs or increased runtime. Our results show that, on multiple circuit topologies with 2,000 test cases, ORACLE reduces runtime by \textbf{20.4x - 104.4x} compared to state-of-the-art approaches. It also meets 99.9\% of the 2,000 target specifications, and achieves \textbf{5.1x - 318.6x} better figure of merit in the resulting output specs.
\end{abstract}


\section{Introduction}
The growing demand for high-performance analog circuits increases the need for automated design flows that reduce manual effort while preserving the accuracy~\cite{fayazi2021applications}. Traditional analog circuit sizing methods generally fall into knowledge-based, model-based, and simulation-based categories, each presenting challenges in adaptability, data efficiency, and computational cost~\cite{liakos2025review,fayazi2023angel,lberni2024ea, fayazi2022fascinet}.

To address these limitations, reinforcement learning (RL) has emerged as a promising approach for analog circuit design by learning device-parameter update policies through repeated interaction with circuit simulations~\cite{uhlmann2022deep,cao2024rose}. This gives RL a key advantage over traditional design methods. Instead of relying on fixed expert heuristics or separately trained surrogate models, an RL agent interacts with the circuit environment, observes the current design state, and applies parameter-update actions. Moreover, the RL agent updates its policy using a reward signal derived from the circuit specifications and design constraints. This closed-loop learning process allows the agent to learn how changes in device parameters affect circuit-level performance, thereby enabling more effective exploration of large and highly constrained analog design spaces~\cite{ahmadzadeh2024using}. 

Despite these advances, most current RL-based analog circuit design automation approaches only consider single-objective optimization functions~\cite{kim2025ppaas,hayes2022practical}. This drastically limits their
applications as circuit design problems usually are multi-objective (MO) optimization problems~\cite{somayaji2024pareto}. Even approaches designed for MO often reduce multiple design specifications to a single scalar reward~\cite{settaluri2020autockt}. Although such approaches simplify training, they hide the true Pareto trade-offs among competing objectives and can therefore lead to suboptimal designs. In addition, because the optimization is driven by a single scalar reward, changes in the desired multi-objective specifications typically require retraining the model from scratch, which is a major limitation for practical analog circuit design automation.automation~\cite{hayes2022practical}.

A common problem in MO-RL is the use of fixed-weight scalarization, where multiple objectives are compressed into a single reward using predefined weights. When objectives have different numerical values, those with larger values can dominate the optimization process, even if they are not the most important under the desired preference. ~\cite{hayes2023brief, basaklar2023pdmorl}. A more suitable alternative is to keep the reward as a vector and learn policies that can recover multiple non-dominated solutions instead of a single weighted-sum solution~\cite{ns2024pareto}. Preferences are still often specified by a weight vector, but a simple weighted sum can remain biased toward high-magnitude objectives. Cosine-aligned guidance addresses this issue by favoring actions whose predicted value vector points in the same direction as the preference vector~\cite{basaklar2023pdmorl}. In other words, it selects actions that improve the objectives in a way that better matches the desired trade-off, rather than simply choosing the action with the largest scalarized value.

In this work, we propose ORACLE, an open-source\footnote{\url{https://anonymous.4open.science/r/ORACLE-2026/README.md}} RL-based multi-objective analog circuit design optimizer approach guided by a large language model (LLM) agent. ORACLE replaces scalar-reward optimization with vector-valued learning, where each design objective has an associated reward signal and the reward is represented as a vector rather than a single scalar value. Consistent with this setting, ORACLE adopts a multi-objective version of Q-learning in which the network learns vectorized Q-values instead of standard scalarized value updates, thereby enabling explicit modeling of trade-offs across specifications. 

The goals of ORACLE are threefold: (a) produce multiple trade-off solutions for each target spec rather than favoring a single reward solution; (b) support preference conditioning so that a single trained policy targets different trade-off regions by changing a preference vector without retraining; (c) improve practical efficiency by reducing unproductive exploration in the discrete parameter-update space. 

To enable preference conditioning, ORACLE uses two preference-guidance mechanisms. The first is normalized weight guidance, which uses preference-weighted action selection over normalized objectives. The second is cosine-aligned guidance, which adds a cosine-similarity term so that the agent prefers actions that are not only numerically large in value, but also point in the same direction as the desired preference trade-off~\cite{basaklar2023pdmorl}.

To improve efficiency, beyond using value guidance, ORACLE applies LLM-guided action masking to add domain knowledge to exploration of the discrete design space. The action masking in discrete RL improves exploration efficiency by filtering actions that are unlikely to be beneficial in the current design state~\cite{cao2025roseopt,wang2024statespecific}. This helps the RL agent make more informed step-by-step parameter changes.

We validate ORACLE on several circuit topologies with 2,000 test cases and compare it against state-of-the-art (SOTA) techniques. Experimental results show that ORACLE meets \textbf{99.9\%} of the target specifications. ORACLE also achieves \textbf{20.4x - 104.4x} runtime speedup. Furthermore, it delivers \textbf{5.1x - 318.6x} higher figure of merit (FoM) for the resulting output specs. 

The main contributions of this paper are summarized as follows:
\begin{itemize}[left=2pt]
\item Proposing ORACLE, an open-source MO analog circuit design optimizer framework that replaces scalar reward optimization with vector-valued learning and explicit trade-off modeling.

\item Introducing preference controllability through normalized weight and cosine aligned guidance to steer optimization toward different trade-off regions under changing specification priorities.

\item Integrating LLM-guided action masking to reduce unbeneficial exploration steps in discrete parameter update spaces, improving both efficiency and reliability.

\item Testing on a benchmark comprising 2,000 target specifications of multiple topologies and compared against SOTA techniques ORACLE meets \textbf{99.9\%} of the specifications, achieves up to a \textbf{104.4x} runtime speedup, and delivers up to a \textbf{318.6x} higher FoM for the resulting designs.
\end{itemize}

\section{Background \& Related Work}

\subsection{Analog Circuit Design Optimization}
Most analog circuit design problems are naturally multiobjective as the desired specifications, \textit{e.g.} gain, bandwidth, etc., should be optimized simultaneously~\cite{li2024robust}. In other words, the goal of analog circuit optimization is to determine the design parameters (\textit{e.g.} DC bias voltages and size of transistors) such that
\begin{equation}
\begin{aligned}
\text{minimize} \quad & f_1(x), \ldots, f_M(x) \\
\text{subject to:} \quad & c_1(x) < 0, \ldots, c_N(x) < 0,
\end{aligned}
\label{eq:analog_opt}
\end{equation}
where $f_1, \ldots, f_M$ are the desired specifications of the circuit~\cite{Fayazi2023FuNToM}. $c_1, \ldots, c_N$ are constraints such as $x_j \in [p_j^-, p_j^+]$ or bandwidth (BW) $> 1\mathrm{GHz}$.

\subsection{Analog Circuit Design Automation}
Previous work on automating analog circuit optimization includes knowledge-based, equation-based, simulation-based, and learning-based methods. They either suffer from generalization across different topologies or are sample-inefficient, which requires restarting the search when the target specification changes~\cite{barros2010analog,daems2002efficient,cohen2015genetic,settaluri2020autockt}. Recently, RL-based methods have improved search efficiency by learning parameter-update policies directly from circuit simulation~\cite{vanhasselt2016double,basaklar2023pdmorl}. 
However, many of them still use a single scalar reward, which combines all design objectives into one score and leads to only one fixed trade-off solution~\cite{settaluri2020autockt,uhlmann2022deep,cao2025roseopt}.

MO-RL methods learn policies across multiple preference settings instead of optimizing a single scalarized objective. Preference-conditioned approaches, such as dynamic-weight, Envelope-based, and Preference-Driven MO RL (PD-MORL) methods, use the preference vector to guide policy or value learning with a single network~\cite{abels2019dynamic,yang2019envelope,basaklar2023pdmorl}. 
Other recent MO-RL methods, including Pareto set learning and decomposition-based approaches, further improve Pareto coverage by explicitly modeling multiple preference directions or subproblems~\cite{felten2024morl,liu2025pareto,hu2024pa2dmorl}. However, in analog circuit design, it is still difficult to efficiently learn many diverse trade-off solutions because the design space is complex and each evaluation requires costly simulation. Similarly, knowledge-guided exploration has been studied to improve RL sample efficiency by filtering clearly unpromising actions through masking or other structured guidance~\cite{cao2025roseopt,wang2024statespecific}. Furthermore, LLMs have shown promising results in analog circuit design automation~\cite{shen2025atelier,lai2025analogcoder,abbineni2026muallm,aldowaish2026sina}. 

\section{Proposed Approach}
\subsection{Overview}
\begin{figure}[t]
\centering
\includegraphics[width=1\columnwidth]{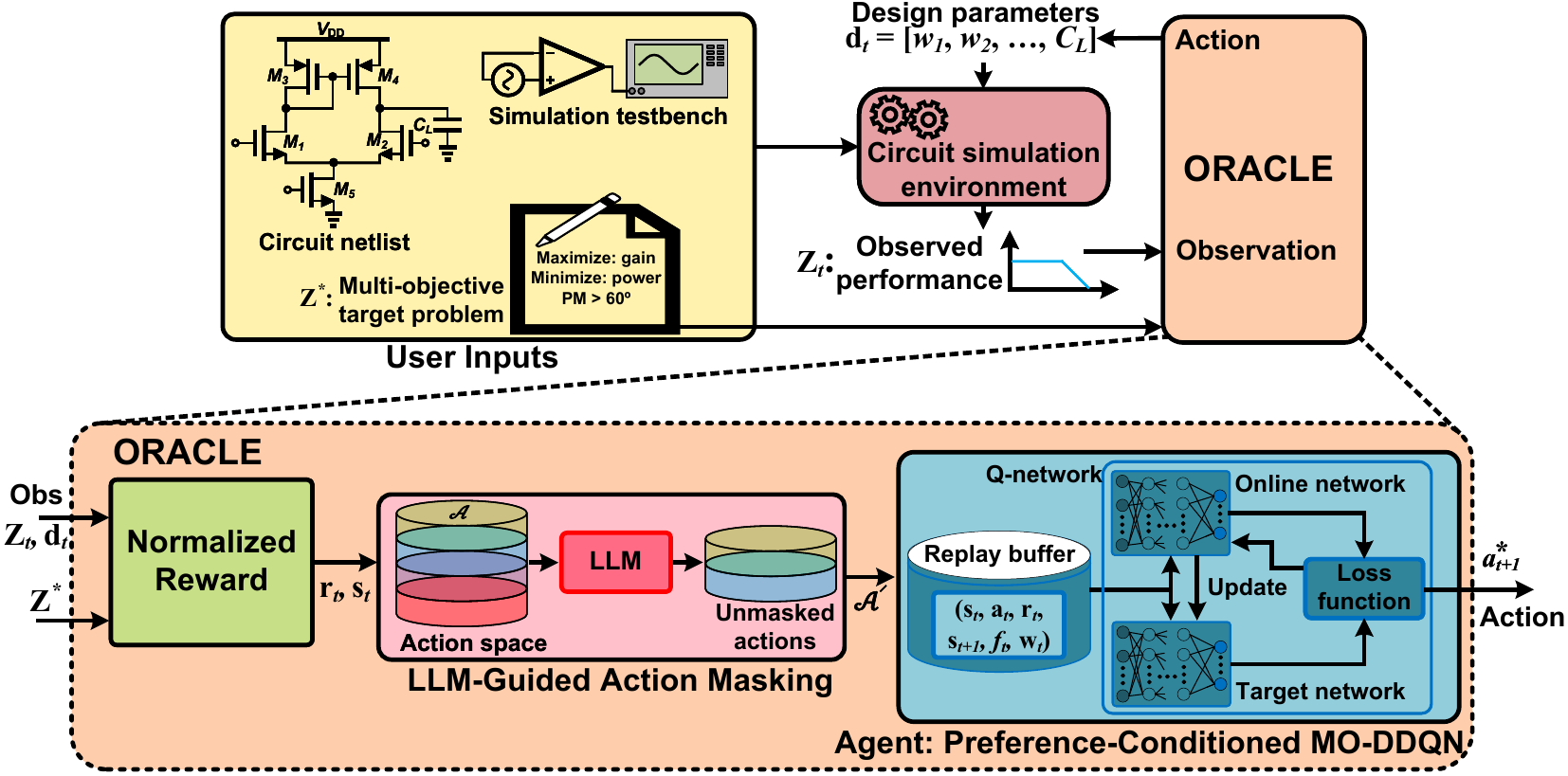}
\caption{Overall ORACLE framework.}
\label{fig:framework}
\end{figure}
ORACLE is a preference-conditioned MO-RL framework for analog circuit design optimization. The goal is to replace the scalarized reward used in conventional RL-based sizing with a vector-valued reward during learning. Instead of training a separate policy for each predefined objective preference, ORACLE conditions a single agent on a preference vector, allowing one trained model to generate multiple trade-off solutions for the same target specification. This addresses a key limitation of scalar-reward RL, which typically produces only one design solution for a given training setup.

The desired circuit specifications is formulated as a MO Markov decision process~\cite{rezaei2025taxonomy}. As shown in Fig.~\ref{fig:framework}, the framework of ORACLE has three main modules: (a) the environment as a normalized reward vector that measures per-objective progress toward the target specification; (b) an LLM-guided masking module to filter undesirable actions before simulation to reduce wasted exploration and improve search efficiency; and (c) a preference-conditioned MO-double deep Q-network (MO-DDQN) to learn an MO action-value function and uses the preference vector for action selection.

\subsection{MO Environment and Normalized Reward}
ORACLE uses a simulator-based circuit environment in which the agent iteratively updates circuit design parameters and observes the resulting circuit specifications, \textit{e.g.,} gain, phase margin, etc. At each step, the agent receives the current circuit state, selects a discrete parameter-adjustment action, runs a circuit simulation, and obtains feedback. Concretely, the action is a tuple of discrete choices, one per design parameter: \{``decrease", ``no change", ``increase"\}. Thus, for each parameter the agent can decrease the current parameter by one unit, apply no change, or increase the parameter by two units; the updated parameters are then clipped to valid bounds before simulation.

At environment step $t$, the state $\mathbf{s}_t$ includes the current observed circuit specifications $\mathbf{z}_t$, the target specifications $\mathbf{z}^{*}$, and the current circuit design parameter values  $\mathbf{d}_t$ (\textit{e.g.,} size of transistors or capacitors value). 
\begin{equation}
\begin{aligned}
\mathbf{s}_t & =
\begin{bmatrix}
\mathbf{z}_t, \mathbf{z}^{*}, \mathbf{d}_t
\end{bmatrix},\\
\mathbf{z}_t & =
\begin{bmatrix}
\mathrm{spec}_t^{(1)}, \ldots, \mathrm{spec}_t^{(N)}
\end{bmatrix},\\
\mathbf{z}^{*} & =
\begin{bmatrix}
{\mathrm{spec}^{(1)}}^{*}, \ldots, {\mathrm{spec}^{(N)}}^{*}
\end{bmatrix},\\
\mathbf{d}_t & =
\begin{bmatrix}
\mathrm{param}_t^{(1)}, \ldots, \mathrm{param}_t^{(M)}
\end{bmatrix},
\end{aligned}
\end{equation}
where $N$ and $M$ are the number of target specifications and the number of circuit design parameters, respectively. 

To prevent objectives with different scales from dominating the reward, the environment returns a normalized $N$-dimensional reward vector ($\mathbf{r}_t$):
\begin{equation}
\mathbf{r}_t=
\begin{bmatrix}
r_t^{\mathrm{spec}^{(1)}}, \ldots, r_t^{\mathrm{spec}^{(N)}}
\end{bmatrix}^{\top}.
\label{eq:reward_vector}
\end{equation}
Each element in this vector represents the relative difference between one objective and its target specification:
\begin{equation}
r_t^{\mathrm{spec}^{(i)}}=\frac{{\mathrm{spec}^{(i)}}_t-{\mathrm{spec}^{(i)}}^{*}}{|{\mathrm{spec}^{(i)}}^{*}|+\epsilon},
\label{eq:reward_pm_ibias}
\end{equation}
where ${spec^{(i)}}^{*}$ denotes the target specification $i$, and $\epsilon > 0$ is a small positive constant introduced to avoid instability. For specifications to be maximized, a positive reward indicates progress toward or beyond the target. Conversely, for specifications to be minimized, the reward term is multiplied by $-1$ so that lower values yield higher rewards. Consequently, all objectives are expressed in a unified maximization framework.

This reward design keeps the objective terms separate during learning and reduces the effect of scale differences across objectives. As a result, the agent identifies whether an action improves each spec, instead of relying on a single combined reward that may hide these differences.


\subsection{Preference-Conditioned MO-DDQN}
The main learning module in ORACLE is a preference-conditioned MO-DDQN. The Q-network ($\mathbf{Q}_{\theta}$) takes a weight vector ($\boldsymbol\omega$) to prioritize the specs and the circuit state ($s$) as inputs and outputs Q-values for each action ($a$), where $\mathbf{Q}_{\theta}(s,a,\boldsymbol{\omega})\in\mathbb{R}^{N}$~\cite{vanhasselt2016double}. In ORACLE, the current circuit state and weight vector are combined and passed through a multilayer perceptron, and the output predicts the effect of each action on each design target. 

To select an action, the predicted vector Q-values are converted into a scalar score using cosine-based similarity with the preference vector \(\boldsymbol{\omega}\)~\cite{basaklar2023pdmorl}:
\begin{equation}
Q_{\mathrm{cos}}(s,a,\boldsymbol{\omega})
=
\cos\!\big(Q_{\theta}(s,a,\boldsymbol{\omega}),\boldsymbol{\omega}\big)\;
\left\|Q_{\theta}(s,a,\boldsymbol{\omega})\right\|_2,
\label{eq:cos_scalarization}
\end{equation}
where \(\cos(\cdot,\cdot)\) measures the directional alignment between the predicted Q-vector and the preference vector, and \(\|\cdot\|_2\) denotes the Euclidean norm. Thus, the cosine term favors actions aligned with the desired trade-off direction, while the norm term favors actions with high Q-values. 

Training follows the DDQN framework with an online Q-network \(Q_{\theta}\), a target Q-network \(Q_{\bar{\theta}}\), and a replay buffer \(\mathcal{D}\). The two networks share the same architecture, but the online network is updated by gradient descent at each training step, whereas the target network is updated less frequently to provide a more stable reference for Q-value updates~\cite{mnih2015dqn,vanhasselt2016doubleq}. The replay buffer stores past transitions and enables random mini-batch sampling, which improves stability by reducing temporal correlation in the training data.
\paragraph{Online and target networks}
The online network \(Q_{\theta}\) is used for action selection. Given the current state \(\mathbf{s}_t\) and preference vector \(\mathbf{w}_t\), it predicts vector Q-values for all discrete actions,
\[
Q_{\theta}(\mathbf{s}_t,\mathbf{w}_t) \in \mathbb{R}^{|\mathcal{A}|\times 4},
\]
and the action is selected by scalarizing these vector Q-values with \(\mathbf{w}_t\) and applying \(\arg\max\). The target network \(Q_{\bar{\theta}}\) is used only during training to compute the next-state target~\cite{vanhasselt2016doubleq}

\paragraph{ Replay buffer}
The replay buffer stores transitions
\[
e_t=\big(\mathbf{s}_t,\; a_t,\; \mathbf{r}_t,\; \mathbf{s}_{t+1},\; f_t,\; \mathbf{w}_t\big),
\]
where \(f_t\in\{0,1\}\) is the terminal flag, and \(\mathbf{w}_t\in\mathbb{R}^{4}\) is the preference vector. During training, mini-batches are uniformly sampled from \(\mathcal{D}\) to improve data reuse and reduce temporal correlation. Storing \(\mathbf{w}_t\) ensures consistency with the preference-conditioned Q-value update.

\paragraph{ DDQN target computation}
For each sampled transition \((\mathbf{s}_t, a_t, \mathbf{r}_t, \mathbf{s}_{t+1}, f_t, \mathbf{w}_t)\), the next action is selected by the online network and then evaluated by the target network, consistent with the DDQN framework~\cite{mnih2015dqn}. The best next action is
\begin{equation}
a^{*}_{t+1}
=
\arg\max_{a'\in\mathcal{A}}
\text{Scalarize}\!\left(Q_{\theta}(\mathbf{s}_{t+1},a',\mathbf{w}_t),\,\mathbf{w}_t\right),
\label{eq:best_action_general_impl}
\end{equation}
and the vector-valued DDQN target is
\begin{equation}
\mathbf{y}_t
=
\mathbf{r}_t
+
\gamma(1-f_t)\,
Q_{\bar{\theta}}\!\left(\mathbf{s}_{t+1}, a^{*}_{t+1}, \mathbf{w}_t\right),
\label{eq:ddqn_target_vec_impl}
\end{equation}
where \(\gamma\in[0,1]\) is the discount factor~\cite{hafner2025worldmodels}.

\paragraph{ Training update}
At each gradient step, the mean-squared error (MSE) loss is computed between the online prediction for the executed action and the target Q-vector,
\[
\mathcal{L}(\theta)=
\left\|
Q_{\theta}(\mathbf{s}_t,a_t,\mathbf{w}_t)-\mathbf{y}_t
\right\|_2^2.
\]
The online parameters \(\theta\) are optimized by backpropagation to minimize \(\mathcal{L}(\theta)\). Every \(K\) steps, the target network parameters are set equal to those of the online network:
\[
\bar{\theta} \leftarrow \theta.
\]
Updating \(Q_{\bar{\theta}}\) less frequently provides more stable targets during learning.

\paragraph{ Cosine and Normalized Weighted (NW) scalarization}
In ORACLE, \(\text{Scalarize}(\cdot,\mathbf{w})\) uses cosine similarity to rank actions under the preference vector \(\mathbf{w}\). This favors actions whose predicted Q-vectors align with the desired trade-off direction. However, because cosine similarity primarily reflects directional agreement rather than improvement magnitude, it may become less discriminative when objective scales differ. In such cases, actions with similar directions can receive similar scores even when their absolute gains are substantially different.

In NW, the same DDQN update is used, but the scalarization in Eq.~\eqref{eq:best_action_general_impl} is replaced by a weighted-sum/NW score:
\[
Q_{\text{NW}}(\mathbf{s},a,\mathbf{w}) \;=\; \mathbf{w}^{\top} Q(\mathbf{s},a,\mathbf{w}).
\]
This addresses the limitation of cosine-based selection by considering not only the direction of the predicted Q-vector, but also the preference-weighted size of its values. NW changes only the action-selection step. The target computation in Eq.~\eqref{eq:ddqn_target_vec_impl} and the rest of the replay-based DDQN training procedure are the same as in Cosine. 

The loss is defined as
\begin{equation}
\mathcal{L}(\theta)
=
\mathbb{E}_{(s_t,a_t,\mathbf{r}_t,s_{t+1},f_t)}
\left[
\left\|
\mathbf{Q}_{\theta}(s_t,a_t,\boldsymbol{\omega})
-
\mathbf{y}_t
\right\|_2^2
\right].
\label{eq:vector_loss}
\end{equation}
This loss measures the mean-squared error between the Q-vector predicted by the online network and the target Q-vector over sampled transitions. A smaller loss means the prediction is closer to the target. Therefore, minimizing \(\mathcal{L}(\theta)\) trains the online network to produce more accurate Q-vector estimates.

The advantage of this design is that a single trained model can generate different trade-off solutions for the same target specification by changing only the preference vector during inference. In ORACLE, we instantiate this capability using a fixed set of 10 preference vectors per target specification, so that it produces 10 solutions for each target without retraining separate policies.

\begin{figure}[t]
\centering
\includegraphics[width=0.8\columnwidth]{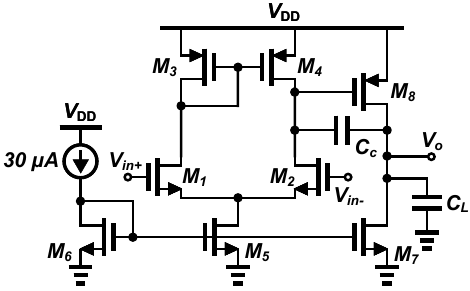}
\caption{Two-stage OPAMP schematic.}
\label{fig:schematic_2stage}
\end{figure}

\begin{figure}[t]
\centering
\includegraphics[width=1\columnwidth]{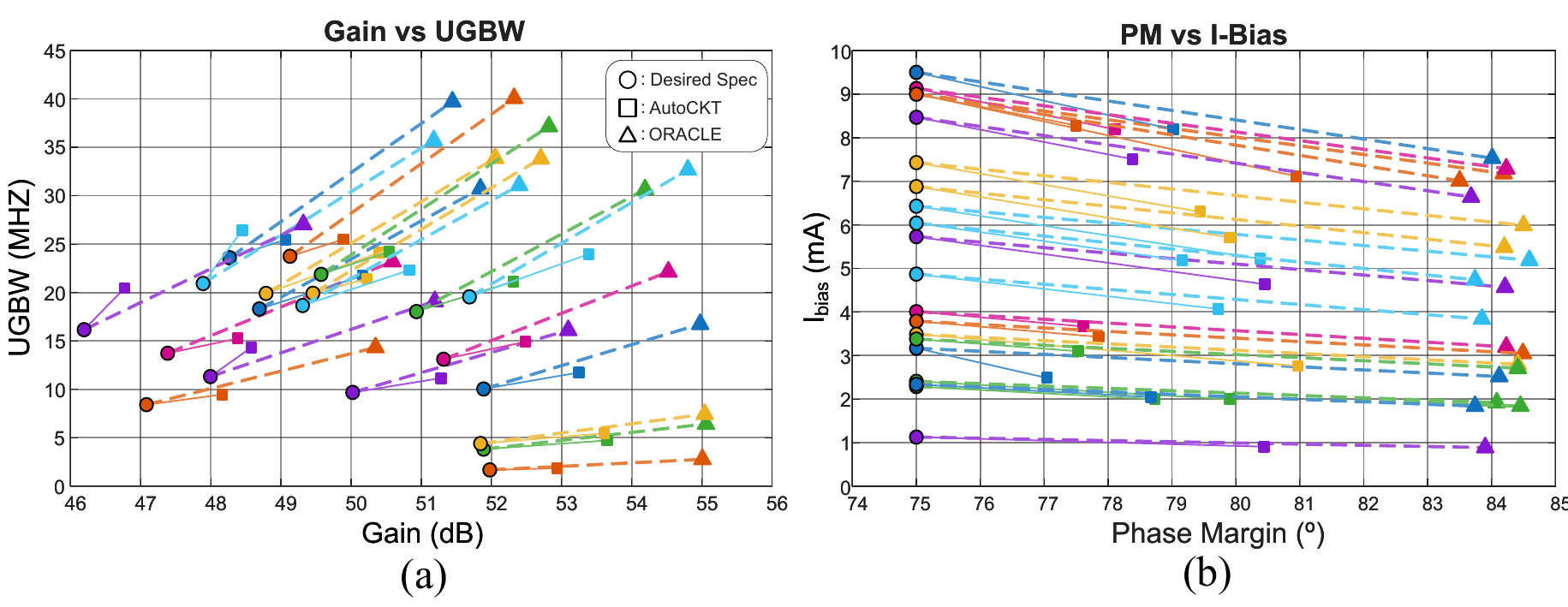}
\caption{Comparison of 20 randomly sampled target specifications from the 1,000 MO problem two-stage OPAMP benchmark and the corresponding results from AutoCkt~\cite{settaluri2020autockt} and ORACLE.} 
\label{fig:random_sol_level_results}
\end{figure}

\subsection{LLM-Guided Action Masking}
Analog circuit simulation is computationally expensive, so reducing unproductive exploration is important during training. To improve exploration efficiency in the discrete design space, ORACLE incorporates an LLM-guided action masking module. At each step, ORACLE compares the current simulated specifications with the target objectives by computing the specification gap
\begin{equation}
\Delta_t^{(i)} = \mathrm{spec}_t^{i} - {\mathrm{spec}^{i}}^{*},
\label{eq:spec_gap}
\end{equation}
where \(\Delta_t^{(i)}\) denotes the deviation of objective \(i\) from its target at step \(t\). These gaps are discretized into four qualitative categories, such as \emph{``far below target"}, \emph{``slightly below target"}, \emph{``near target"}, and \emph{``meets target"}, and then are used to prompt a local LLM.

Using this summarized objective-gap description, the LLM applies circuit-design knowledge to identify parameter changes that are unlikely to move the circuit closer to the target specifications. For example, it may block transistor upsizing when \(I_{\text{bias}}\) exceeds its target, or block downsizing when gain remains far below its target. In this way, globally harmful actions are filtered before simulation.

The LLM output is converted into a binary mask $\mathbf{m}_t \in \{0,1\}^{|\mathcal{A}|}$, where \(m_t(a)=0\) indicates that action \(a\) is filtered out and \(m_t(a)=1\) indicates that action \(a\) is retained. Action selection is then restricted to the unmasked subset,
\begin{equation}
\begin{aligned}
\mathcal{A'}&=\{a\in\mathcal{A}\mid m_t(a)=1\}\\
a_t^{*}&=\arg\max_{a\in\mathcal{A'}}
Q_{\text{cos}}(s_t,a,\boldsymbol{\omega}),
\label{eq:masked_action_selection}
\end{aligned}
\end{equation}
so that the policy searches over a smaller and more relevant subset of actions rather than the full discrete action space. This improves sample efficiency by reducing unnecessary simulation calls.

\section{Evaluation}

\begin{table}[t]
\begin{center}
\begin{threeparttable}
\centering
\caption{Pass-rate and FoM comparison on 1,000 multi-objective
two-stage op-amp problems.}
\def\arraystretch{1.1}
\label{tab:overall_results}
\begin{tabular}{|M{31mm}|M{12mm}|M{13mm}|M{15mm}|}
\hline\hline
\textbf{Method} &
\textbf{Pass-rate} &
\textbf{Average FoM} &
\textbf{Top-20 FoM} \\
\hline
MODEBI~\cite{vicsan2022automated} & 9.0\% & 2.00 & 3.56\\
\hline
ABCMOBO~\cite{zhao2024asynchronous} & 75.8\% & 8.50 & 34.07\\
\hline
AutoCKT~\cite{settaluri2020autockt} & 93.8\% & 0.43 & 0.70 \\
\hline
ORACLE (Cosine) & 100.0\% & 1.45 & 1.49\\
\hline
ORACLE (Cosine + LLM) & 100.0\% & 132.3 & 384.3\\
\hline
\mycc ORACLE (NW) & \mycc \textbf{100.0\%} & \mycc \textbf{138.3} & \mycc \textbf{450.1} \\
\hline
\end{tabular}
\end{threeparttable}
\end{center}
\end{table}

\begin{figure}[t]
\centering
\includegraphics[width=0.9\columnwidth]{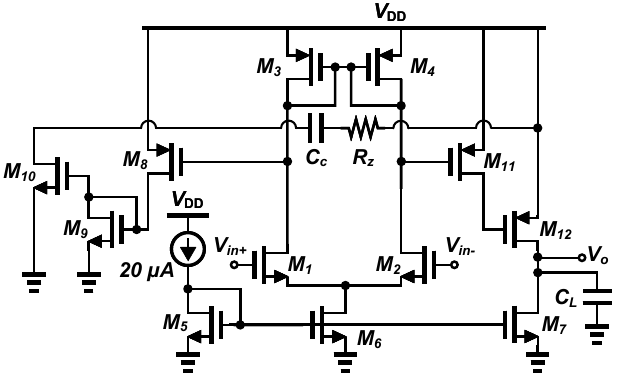}
\caption{Unconventional three-stage OTA schematic.}
\label{fig:schematic_OTA}
\end{figure}

\begin{table}[t]
\centering
\renewcommand{\arraystretch}{1.1}
\caption{Average runtime on the two-stage op-amp benchmark.}
\label{tab:2stage_runtime}
\begin{tabular}{|M{12mm}|M{4mm}|M{4mm}|M{4mm}|M{10mm}|M{15mm}|M{10mm}|}
\hline
\hline
\textbf{Method} & \cite{vicsan2022automated} & \cite{settaluri2020autockt} & \cite{zhao2024asynchronous} & ORACLE (cos) & ORACLE (cos + LLM) & ORACLE (NW)\\
\hline
\textbf{Runtime (Minutes)} & 180 & 85 & 59.9 & 6.5 & 3.2 & \cellcolor{gray!35}\textbf{2.4} \\
\hline
\end{tabular}
\end{table}
We evaluate ORACLE on a two-stage OPAMP and an unconventional 3-stage OTA topologies benchmark in 45nm BSIM technology node as shown in Fig.~\ref{fig:schematic_2stage} and Fig.~\ref{fig:schematic_OTA}. The general desired target optimization problems for both these topologies are as follows: 
\begin{equation}
\begin{aligned}
\text{maximize} \quad & \{G,~UGBW,~PM\} \\
\text{minimize} \quad & \{I_{bias}\} \\
\text{subject to:} \quad & G~\geq~G^*, UGBW~\geq~UGBW^*,\\
\quad & PM~\geq~PM^*, I_{bias}~\leq~I_{bias}^*.
\label{eq:eval_optimization}
\end{aligned}
\end{equation}
Here, $(\cdot)$ and $(\cdot)^*$ denote the achieved specification and corresponding target specification, respectively. A solution is considered successful only if it meets all four target specifications at the same time \textit{i.e.} $G~\geq~G^*$, $UGBW~\geq~UGBW^*$, $PM~\geq~PM^*$, and $I_{bias}~\leq~I_{bias}^*$.

The evaluation set for each topology contains 1,000 optimization problems in~\eqref{eq:eval_optimization} format, where the target specifications are sampled from the ranges specified in \eqref{eq:eval_spec_range} and \eqref{eq:eval_spec_range_ota} for each topology, respectively. For each optimization, AutoCKT~\cite{settaluri2020autockt}, which uses a scalar reward, returns one solution. On the other hand, ORACLE generates 10 solutions (as it is set in its preference settings) for the same target optimization problem. This is because of the multi-objective optimization of ORACLE, which generates multiple trade-off solutions from a single trained model. To have a fair evaluation, we also compare ORACLE with two MO Bayesian optimization methods, MODEBI~\cite{vicsan2022automated} and ABCMOBO~\cite{zhao2024asynchronous}, with the same 10-solution setting. A Figure of Merit (FoM) is defined as follows to rank these 10 solutions and select the best one: 

\begin{equation}
\begin{aligned}
\mathrm{FoM} =\;&
\frac{G-G^{*}}{G^{*}} + \frac{U-U^{*}}{U^{*}} + \frac{PM-PM^{*}}{PM^{*}} + \frac{I^{*}_{\mathrm{bias}}-I_{\mathrm{bias}}}{I^{*}_{\mathrm{bias}}}.
\end{aligned}
\label{eq:fom}
\end{equation}

\begin{table}[t]
\centering
\renewcommand{\arraystretch}{1.1}
\caption{Comparison on 10,000 generated solutions over the same 1,000
multi-objective two-stage OPAMP benchmark, with 10 solutions generated
per target specification.}
\label{tab:ORACLE_10000_results}
\begin{tabular}{|M{28mm}|M{15mm}|M{14mm}|M{14mm}|}
\hline\hline
\textbf{Method} & \textbf{Pass-rate} & \textbf{Average FoM} & \textbf{Top-20 FoM} \\
\hline
MODEBI~\cite{vicsan2022automated} & 1.40\% & 1.89 & 3.77 \\
\hline
ABCMOBO~\cite{zhao2024asynchronous} & 40.37\% & 4.82 & 35.92 \\
\hline
ORACLE (Cosine) & 99.19\% & 107.38 & 355.37 \\
\hline
ORACLE (Cos + LLM) & 99.45\% & \mycc \textbf{113.35} & \mycc \textbf{460.08} \\
\hline
\mycc ORACLE (NW) & \mycc \textbf{99.80\%} & 107.46 & 372.40 \\
\hline
\end{tabular}
\end{table}

\begin{table}[t]
\centering
\caption{Pareto trade-off quality comparison on the two-stage OPAMP. PF: Pareto front.}
\label{tab:pareto_results}
\def\arraystretch{1.05}
\begin{threeparttable}
\begin{tabular}{|M{19mm}|M{19mm}|M{16mm}|M{17mm}|}
\hline\hline
\textbf{Method} & \textbf{Mean Hypervolume} & \textbf{Mean Sparsity} & \textbf{Mean PF Size} \\
\hline
AutoCKT~\cite{settaluri2020autockt} & $1.40\times 10^{5}$ & 0.00 & 1.00 \\
\hline
MODEBI~\cite{vicsan2022automated} & $6.28\times 10^{6}$ & \mycc $\mathbf{2.94\times 10^{2}}$ & \mycc \textbf{10.00} \\
\hline
ABCMOBO~\cite{zhao2024asynchronous} & $3.29\times 10^{6}$ & $4.99\times 10^{2}$ & \mycc \textbf{10.00} \\
\hline
ORACLE (Cos) & $3.55\times 10^{9}$ & $8.81\times 10^{5}$ & \mycc \textbf{10.00} \\
\hline
ORACLE (NW) & \mycc $\mathbf{3.57\times 10^{9}}$ & $9.04\times 10^{5}$ & \mycc \textbf{10.00} \\
\hline
\end{tabular}
\end{threeparttable}
\end{table}

It should be noted that the FoM is computed only over successful solutions, \textit{i.e.}, designs that satisfy all four constraints. For each target specification, the reported FoM corresponds to the highest-FoM successful solution. 

\begin{figure}[t]
\centering
\includegraphics[width=1\columnwidth]{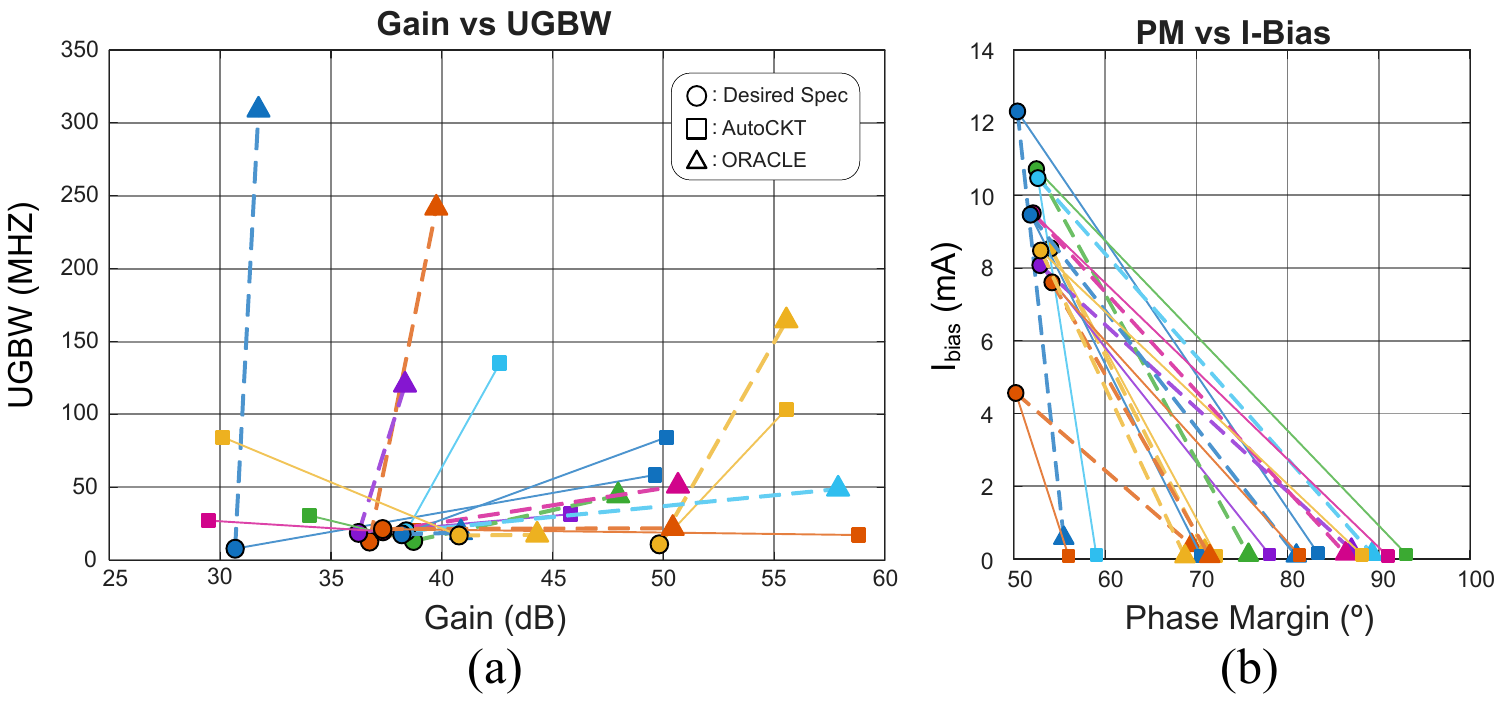}
\caption{Comparison of 10 randomly sampled target specifications from the 1,000 MO problem benchmark of 3-stage unconventional OTA and the corresponding results from AutoCkt~\cite{settaluri2020autockt} and ORACLE.}
\label{fig:random_sol_level_results_ota}
\end{figure}

\begin{figure}[b]
\centering
\includegraphics[width=1\columnwidth]{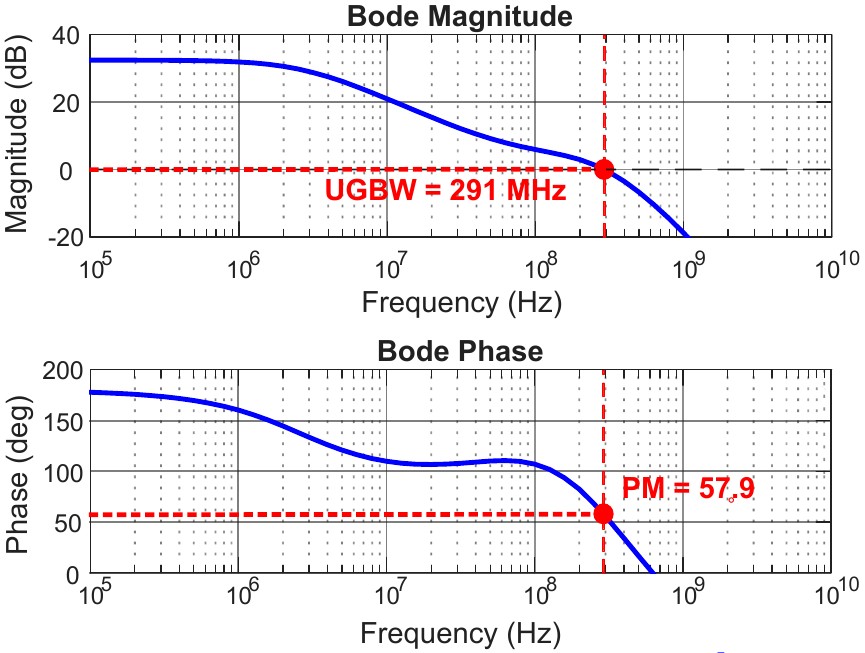}
\caption{The frequency response of the ORACLE-generated circuit for Table~\ref{tab:sample_ota_design} target specifications.}
\label{fig:sample_ota_design}
\end{figure}

We compare the following methods:
\begin{enumerate}
    \item AutoCKT: A single-objective RL optimizer producing one solution per target~\cite{settaluri2020autockt} which is one of our SOTA baselines.

    \item MODEBI: A MO Bayesian optimizer producing 10 solution per target~\cite{vicsan2022automated}, which is one of our SOTA baselines.

    \item ABCMOBO: An asynchronous batch constrained MO Bayesian optimizer producing 10 solution per target~\cite{zhao2024asynchronous}, which is one of our SOTA baselines.
    
    \item ORACLE (Cosine): Preference-conditioned MO-DDQN with cosine-aligned action selection.
    
    \item ORACLE (Cosine + LLM): Cosine-based ORACLE augmented with LLM-guided action masking.
    
    \item ORACLE (NW): ORACLE with normalized weighted-sum scalarization for action selection.
\end{enumerate}

Our evaluation metrics are as follows:
\begin{itemize}
\item Pass-rate: Measures the percentage of the generated designs that satisfy all four target constraints simultaneously

\item FoM~\eqref{eq:fom}: Indicates the design quality. Higher FoM values mean better solutions.

\item Hypervolume~\cite{liu2025pareto}: Measures how much useful objective space is covered by the recovered non-dominated solutions. 

\item Sparsity: Measures how well these solutions are distributed across the trade-off surface. Lower sparsity means the Pareto solutions are distributed more evenly, which is better. 

\item Pareto front size: Measures the number of non-dominated solutions obtained for each target specification. 
\end{itemize}
ORACLE is implemented in Python using PyTorch for the NN components. Training uses a preference-conditioned DDQN algorithm with experience replay and \(\epsilon\)-greedy exploration in the AutoCkt multi-objective OpenAI Gym environment. 
The models are trained on 50 target specifications with 10 preference vectors and a maximum episode length of 30 environment steps. For the LLM-guided method, a local Ollama backend runs a lightweight Llama 3.2 model for action masking and preference-boost recommendations. 
All experiments are conducted on a machine equipped with an NVIDIA GA102 GPU. 

\begin{table}[t]
\centering
\renewcommand{\arraystretch}{1.1}
\caption{Pass-rate and FoM comparison on 1,000 multi-objective
unconventional three-stage OTA problems.}
\label{tab:ota_solution_level}
\begin{tabular}{|M{30mm}|M{15mm}|M{14mm}|M{14mm}|}
\hline\hline
\textbf{Method} & \textbf{Pass-rate} & \textbf{Average FoM} & \textbf{Top-20 FoM} \\
\hline
MODEBI~\cite{vicsan2022automated} & 0.9\% & 0.52 & 3.34 \\
\hline
AutoCKT~\cite{settaluri2020autockt} & 35.7\% & 3.07 & \cellcolor{gray!35}\textbf{49.20} \\
\hline
ABCMOBO~\cite{zhao2024asynchronous} & 70.4\% & 2.61 & 7.54\\ 
\hline
ORACLE (Cosine) & 97.2\% & 8.13 & 29.20 \\
\hline
ORACLE (NW) & 96.4\% & \cellcolor{gray!35}\textbf{13.30} & 40.70 \\
\hline
ORACLE (Cos + LLM) & \cellcolor{gray!35}\textbf{98.4\%} & 7.15 & 35.77 \\
\hline
\end{tabular}
\end{table}

\begin{table}[t]
\centering
\renewcommand{\arraystretch}{1.1}
\caption{A sample of the target specs from the 3-stage unconventional OTA benchmark and the corresponding result of ORACLE.}
\label{tab:sample_ota_design}
\begin{tabular}{|M{12mm}|M{12mm}|M{18mm}|M{10mm}|M{12mm}|}
\hline\hline
\textbf{Specs} & Gain (dB) & UGBW (MHz) & PM ($^\circ$) & $I_{\text{bias}}$ (mA)\\
\hline
\textbf{Target} & $>$ 31.7 & $>$ 4.83 & $>$ 50.7 & $<$ 13.1 \\
\hline
\textbf{ORACLE} & 32.4 & 291.6 & 57.9 & 0.53 \\
\hline
\end{tabular}
\end{table}

\subsection{Two-stage OPAMP}
The circuit schematic is shown in Fig.~\ref{fig:schematic_2stage}. The action space for each circuit component is written in array notation: [start, end, increment]. The action space for every transistor width is [1, 100, 1]$\times0.5\mu$. Note that the size of $M_1=M_2$ and the size of $M_3=M_4$. Moreover, the action space of the compensation capacitor ($C_c$) is [0.1, 10.0, 0.1]$\times$1pF. This results in 7 design parameters with a total action space size of $10^{14}$ possible values. The design specifications of interest and their range in our database are as follows: 

\begin{equation}
\begin{aligned}
\text{Gain}: \; [46dB, 52dB]~\mathrm{V/V}, & \;
PM :\; [75^\circ], \\
UGBW :\; [1, 25]~\mathrm{MHz}, & \;
I_{\text{bias}} :\; [0.1, 10]~\mathrm{mA}.
\end{aligned}
\label{eq:eval_spec_range}
\end{equation}

Table~\ref{tab:overall_results} summarizes the overall comparison. ORACLE meets \textbf{100\%} of the target specifications, representing a \textbf{1,011\% - 6.6\%} improvement over the SOTA~\cite{settaluri2020autockt, vicsan2022automated, zhao2024asynchronous}. Moreover, ORACLE (NW) delivers \textbf{318.6x - 69.15x} improvement in FoM compared to SOTA~\cite{settaluri2020autockt, vicsan2022automated, zhao2024asynchronous}. Also, comparing ORACLE (Cosine) and ORACLE (Cosine + LLM) demonstrates a 91x improvement on average FoM, indicating the positive effect of LLM integration. Furthermore, we compare the average FoM of top twenty best designs of each method. ORACLE (NW) achieves 302x - 1.17x better top-20 FoM compared to ORACLE (Cosine) and ORACLE (Cosine + LLM), respectively. This shows the effectiveness of the proposed normalized weighted-sum scalarization for action selection. 

Fig.~\ref{fig:random_sol_level_results} shows a comparison between the output of ORACLE and AutoCKT over 20 randomly sampled target specifications from the 1,000 MO problem benchmark. As illustrated, across all samples, ORACLE achieves higher gain, UGBW, and phase margin than AutoCkt, while maintaining lower $I_{\text{bias}}$. Moreover, as shown in Table~\ref{tab:2stage_runtime}, ORACLE reduces the average runtime by \textbf{75x - 24.8x} compared with SOTA~\cite{settaluri2020autockt, vicsan2022automated, zhao2024asynchronous}. 

We also evaluate the results of all 10,000 generated solutions. The test is still on the same 1,000 MO problem benchmark, but now we keep all 10 generated solutions for each MO target problem. The results are listed in Table~\ref{tab:ORACLE_10000_results}. The NW-based method achieves better performance than the cosine-based approach in all aspects. 
This suggests that, in the implementation, direct weighted scoring leads to more effective action selection and therefore improves the ability to find valid solutions. Moreover, the LLM-guided masking module reduces wasted exploration and results in more passed solutions. 

\begin{table}[t]
\centering
\renewcommand{\arraystretch}{1.1}
\caption{Average unconventional three-stage OTA runtime comparison.}
\label{tab:ota_runtime}
\begin{tabular}{|M{12mm}|M{4mm}|M{4mm}|M{4mm}|M{10mm}|M{15.5mm}|M{10mm}|}
\hline\hline
\textbf{Method} &
\cite{settaluri2020autockt} & \cite{vicsan2022automated} & \cite{zhao2024asynchronous} & ORACLE (Cos) & ORACLE (Cos + LLM) &
ORACLE (NW)\\
\hline
\textbf{Runtime (Minutes)} &
338 & 295 & 66.1 & 10.54 & 11.58 & \cellcolor{gray!35}\textbf{3.24} \\
\hline
\end{tabular}
\end{table}

\begin{table}[t]
\centering
\renewcommand{\arraystretch}{1.1}
\caption{Pareto trade-off quality comparison on the Current Mirror OTA.
PF denotes Pareto front.}
\label{tab:ota_pareto}
\begin{tabular}{|M{28mm}|M{18mm}|M{13mm}|M{12mm}|}
\hline\hline
\textbf{Method} &
\textbf{Mean Hypervolume} &
\textbf{Mean Sparsity} &
\textbf{Mean PF Size} \\
\hline
AutoCKT~\cite{settaluri2020autockt} & $6.12 \times 10^{6}$ & 0 & 1 \\
\hline
MODEBI~\cite{vicsan2022automated} & $9.91 \times 10^{6}$ & \textbf{$9.46 \times 10^{5}$} & \cellcolor{gray!35}\textbf{10} \\
\hline
ABCMOBO~\cite{zhao2024asynchronous} & $1.91 \times 10^{7}$ & $4.96 \times 10^{5}$ & \cellcolor{gray!35}\textbf{10}\\
\hline
ORACLE (NW) & \cellcolor{gray!35}$\mathbf{3.54 \times 10^{7}}$ & $2.61 \times 10^{5}$ & \cellcolor{gray!35}\textbf{10} \\
\hline
ORACLE (Cos) & $1.32 \times 10^{7}$ & \cellcolor{gray!35}$\mathbf{2.17 \times 10^{5}}$ & \cellcolor{gray!35}\textbf{10} \\
\hline
ORACLE (Cos + LLM) & $1.60 \times 10^{7}$ & $3.10 \times 10^{5}$ & \cellcolor{gray!35}\textbf{10} \\
\hline
\end{tabular}
\end{table}

Table~\ref{tab:pareto_results} compares the Pareto-quality metrics. 
ORACLE (NW) produces meaningful trade-off sets, with the highest mean hypervolume compared to all other approaches. This shows that the ORACLE reward formulation improves more than one operating point; it yields a substantially richer Pareto front for each target specification. 

\subsection{Unconventional Three-stage OTA}
%

The circuit schematic is shown in Fig.~\ref{fig:schematic_OTA}. The action
space for every transistor width is
[4, 40, 2]$\times0.5\mu$. Moreover, the action space of the compensation capacitor ($C_c$) is [0.05, 2.0, 0.05]$\times$1pF and of the zero-nulling resistor ($R_z$) is [0.1, 10.0, 0.1]$\times$1k$\Omega$. This results in 10 design parameters with a total action space size of $10^{12}$ possible values. The design specifications of interest and their range in our database are as follows:


\begin{equation}
\begin{aligned}
\text{Gain}: \; [23dB, 53dB]~\mathrm{V/V}, & \;
PM :\; [45^\circ, 60^\circ], \\
UGBW :\; [4, 33]~\mathrm{MHz}, & \;
I_{\text{bias}} :\; [0.5, 20]~\mathrm{mA}.
\end{aligned}
\label{eq:eval_spec_range_ota}
\end{equation}

Table~\ref{tab:ota_solution_level} reports the overall comparison. ORACLE (Cos + LLM) meets 98.4\% of 1,000 target specifications, which shows \textbf{109.3x - 1.4x} improvement over SOTA~\cite{settaluri2020autockt, vicsan2022automated, zhao2024asynchronous}. In addition, ORACLE (NW)'s average FoM is \textbf{5.11x - 25.58x} higher than SOTA~\cite{settaluri2020autockt, vicsan2022automated, zhao2024asynchronous}. Also, Fig.~\ref{fig:random_sol_level_results_ota} illustrates 10 of these target specifications and the corresponding results from ORACLE and SOTA for different specifications. Table~\ref{tab:sample_ota_design} lists an example of these target specs and the ORACLE achieved results for it. Fig.~\ref{fig:sample_ota_design} demonstrates the frequency response of the ORACLE output using the unconventional 3-stage OTA. Moreover, as shown in Table~\ref{tab:ota_runtime}, ORACLE reduces the average runtime by \textbf{104.4x - 20.4x} compared with SOTA~\cite{settaluri2020autockt, vicsan2022automated, zhao2024asynchronous}.

Table~\ref{tab:ota_pareto} lists the Pareto-front quality metrics. ORACLE (NW) achieves the highest mean hypervolume, and ORACLE (Cosine) achieves the lowest mean sparsity among all.

\section{Conclusion}
This work presents ORACLE, a novel open-source multi-objective RL-based framework for analog circuit design optimization. ORACLE replaces scalar-reward optimization with preference-conditioned vector-valued learning and further integrates LLM-guided exploration. By preserving objective-level information throughout training, ORACLE recovers multiple trade-off solutions for the same target specification from a single trained model. ORACLE is compared with SOTA techniques on a benchmark with 2,000 different MO specifications. The results show that ORACLE meets 99.9\% of the target specifications. Moreover, ORACLE delivers up to a 318.6x higher FoM for the resulting output specs. In addition, ORACLE achieves up to a 104.4x runtime speedup.

\bibliographystyle{IEEEtran}

\end{document}